\documentclass[12pt,preprint]{aastex}

\shorttitle{Differential Rotation}
\shortauthors{Herbst, Dhital, Francis, Lin, Tresser \& Williams}

\begin{document}


\title{Evidence for Differential Rotation on a T Tauri Star}


\author{William Herbst\altaffilmark{1}, Saurav Dhital\altaffilmark{2}, Alice Francis\altaffilmark{3}, LiWei Lin\altaffilmark{4}, Nyla Tresser\altaffilmark{1} \and Eric Williams\altaffilmark{1}}
\affil{Astronomy Department, Wesleyan University, Middletown, CT 06459}




\altaffiltext{1}{Astronomy Dept., Wesleyan University, Middletown, CT 06459}
\email{wherbst@wesleyan.edu, ntresser@wesleyan.edu, ewilliams@wesleyan.edu}

\altaffiltext{2}{Physics and Astronomy Dept., Swarthmore College, 
Swarthmore, PA 19081}
\email{sdhital1@swarthmore.edu}

\altaffiltext{3}{Physics Dept., Hamilton College, Clinton, NY 13323}
\email{afrancis@hamilton.edu}
\altaffiltext{5}{Harvard-Smithsonian Center for Astrophysics, 60 Garden St., Cambridge, MA 02138}
\email{llin@head.cfa.harvard.edu}


\begin{abstract}

Five years of photometric monitoring of the T Tauri star HBC 338 in NGC 1333 has revealed that it is a periodic variable, but the period has changed significantly with time. From 2000-2003, a period near 5.6 days was observed, while in the last two seasons, the dominant period is near 4.6 days. No other T Tauri star has been seen to change its period by such a large percentage. We propose a model in which a differentially rotating  star is seen nearly equator-on and a high latitude spot has gradually been replaced by a low latitude spot. We show that this model provides an excellent fit to the observed shapes of the light curves at each epoch. The amplitude and sense of the inferred differential rotation is similar to what is seen on the Sun. This may be surprising given the likely high degree of magnetic surface activity on the star relative to the Sun but we note that HBC 338 is clearly  an exceptional T Tauri star. 
    
\end{abstract}




\section{Introduction}

T Tauri stars are the earliest optical manifestation of the star formation process for solar-like stars. As such, they deserve intense scrutiny for what they may reveal about the process in general and the history of the Sun and solar system in particular. Interpretation of their spectra and, therefore, derivation of basic properties such as effective temperature, surface gravity, radius, mass and age, is complicated by the fact that their atmospheres are so unlike those of older, quiescent stars. In particular, they are probably saturated with strong magnetic fields \citep{jkvs} and demonstrably replete with zones of much lower or higher than average surface temperature \citep{h94}. These ``spots" and accretion or flare-heated zones are variable on time scales of days or even hours. It will probably remain a challenge for some time to understand even the basics of their atmospheres.

Since their atmospheres, and at least some of their interiors, are highly magnetized these stars may also be expected to show different surface rotation properties than, say, the Sun. In fact, rotation period studies which have extended over more than a decade in some cases, rarely show any variations in rotation period which are larger than the errors of the determinations \citep{chw04} and none exceeding a few percent. The Sun, by contrast, shows a rotation period gradient of about 40\% between its equatorial and polar regions, with the equator spinning faster \citep{b89}. One may infer that T Tauri stars do not, in fact, have much differential rotation on their surfaces or that their spots are mostly confined to a small latitude zone, or both.

It was a surprise, therefore, to discover the anomalous behavior of one T Tauri star among about a thousand monitored regularly at Van Vleck Observatory (VVO) on the campus of Wesleyan University. It is a member of the young cluster NGC 1333 and otherwise unremarkable. We bring attention to it here because it proves, we argue, that differential rotation can occur on the surfaces of T Tauri stars, and that in this one case where it is observed, both the magnitude and sense of the gradient is similar to what is observed on the Sun. It also shows that large spots distributed inhomogeneously in longitude, can exist in the equatorial zones of T Tauri stars. The star is perhaps best thought of as the ``exception that proves the rule". If period changes like this were common among T Tauri stars we would easily have found more of them.

\section{Observations}

CCD images of two slightly overlapping fields in NGC 1333 were obtained over five observing seasons from November 2000 to April 2005  with the 0.6 m Perkin telescope at Van Vleck Observatory (VVO) on the campus of Wesleyan University in Middletown, CT. At each epoch, a series of five one-minute exposures was taken through a Cousins I filter. The five images were shifted and combined in a pipeline reduction to form a single image equivalent to a five minute exposure but with increased dynamic range and better image quality than would otherwise have been obtained. The shifted and combined images were then bias- and dark-corrected and flat-fielded using twilight flats. The CCD is a 1024 x 1024 front-illuminated array with a pixel size of 0.6$\arcsec$. Seeing, as judged by the full width at half maximum of a stellar profile, ranged from 1.5$\arcsec$ to 2.8$\arcsec$ with a median value of $\sim$ 2.5$\arcsec$. Differential aperture photometry was carried out relative to a set of non-variable stars on the same images. Full results from this study will be presented elsewhere and additional description of our methods is given by \citet{chw04}.

\begin{figure}
\plotone{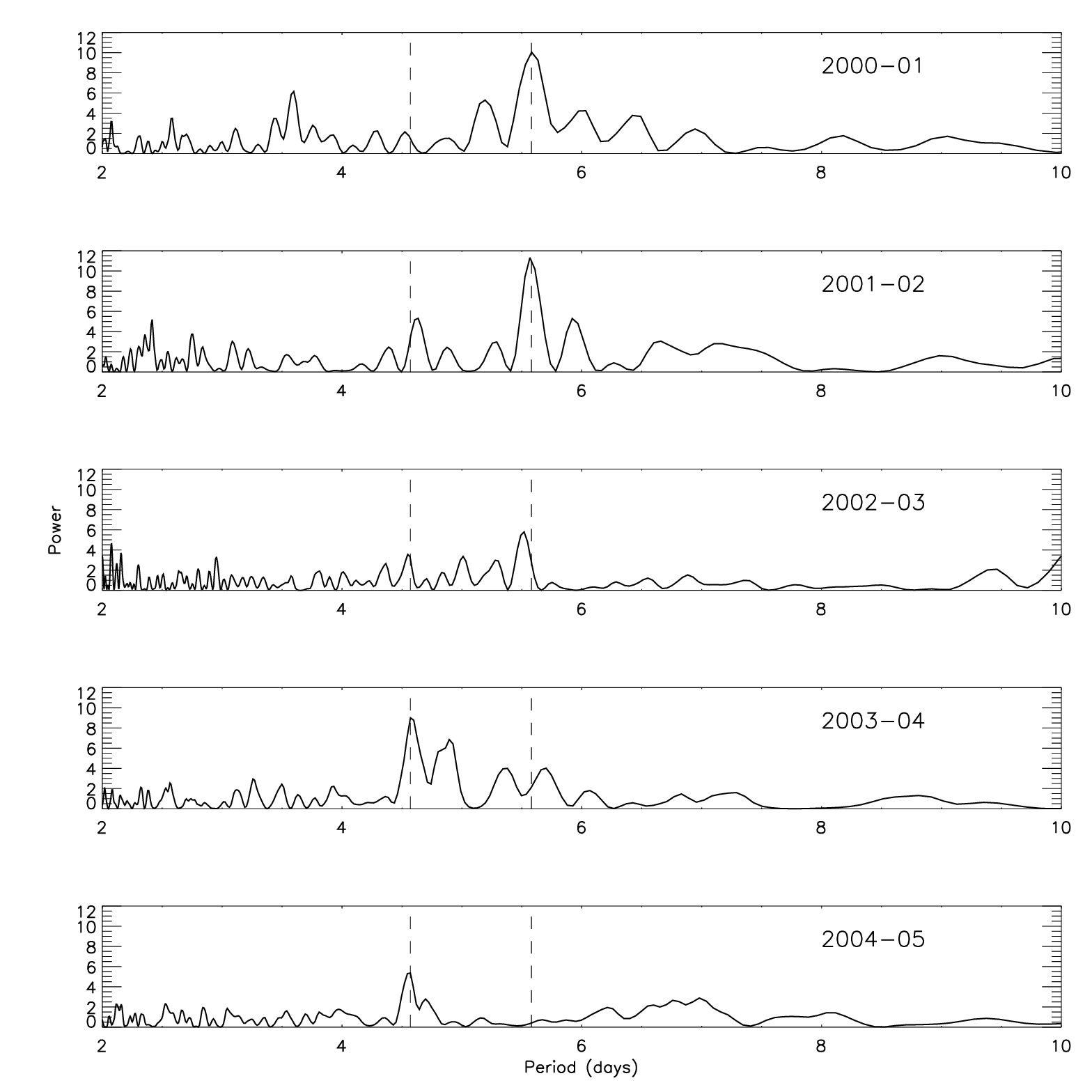}
\caption{Periodograms calculated by the Lomb-Scargle technique of the data on HBC 338 obtained in each of five seasons at VVO. It is clear that in the first two seasons the dominant period was near 5.6 days while in the last two years it was near 4.6 days. Some power near both periods was seen in the transition year, 2002/03.} 
\label{periodograms}
\end{figure}

\begin{figure}
\plotone{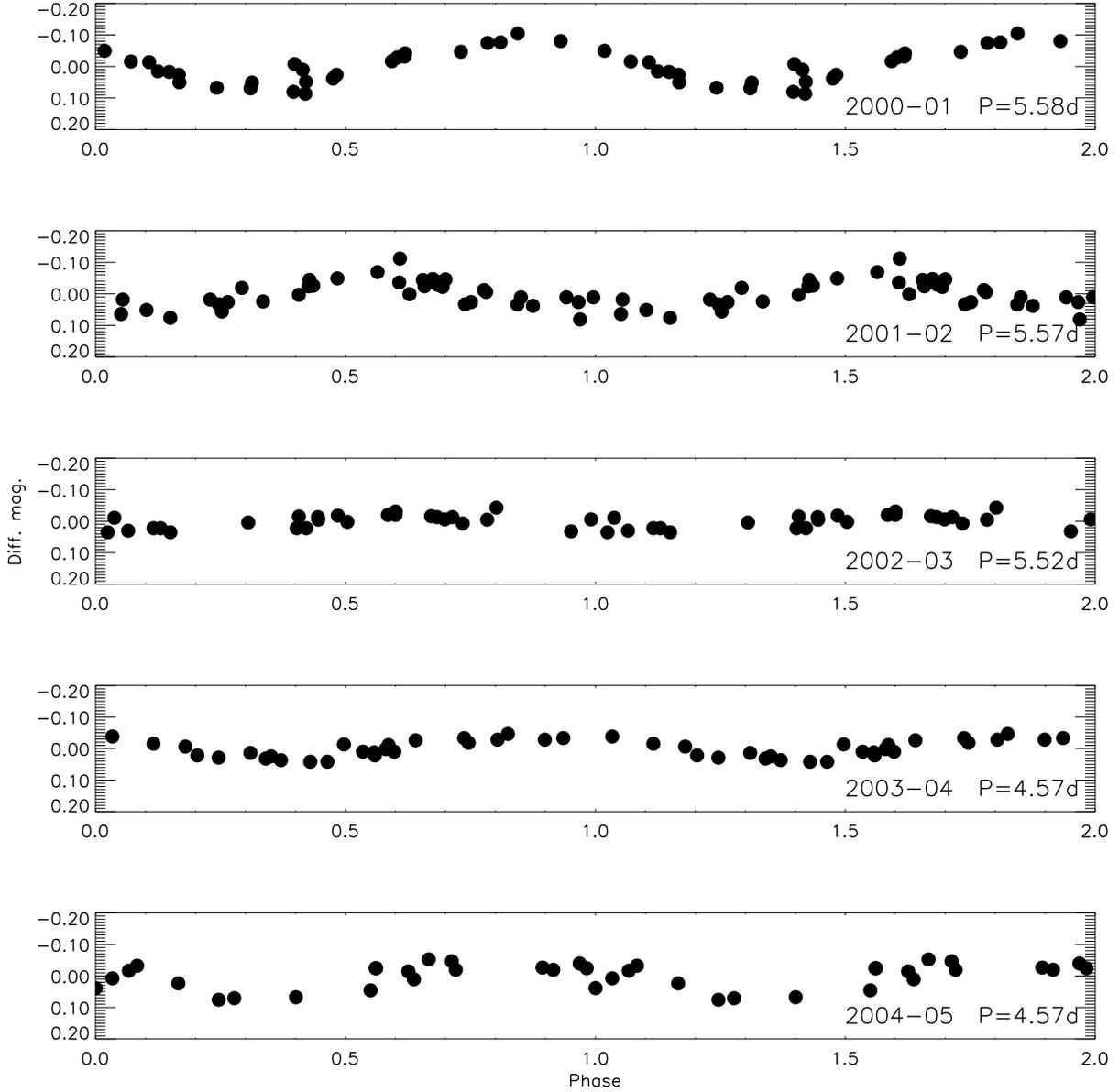}
\caption{Phased light curves for each season based on the period with the highest power. The evolution from a more continuously variable shape in the first two seasons, with a peaked maximum, to a  flatter maximum in the last two seasons is apparent. The increased noise in the third season reflects the presence of more than one spot with more than one period, as revealed by the periodogram.} 
\label{lightcurves}
\end{figure}

Our data were searched for periodicity using the Lomb-Scargle periodogram technique described, for example, by \citet{hb86}. A number of periodic variables were discovered and, as is the case in other clusters, when a star was found to be periodic in more than one season, the periods always agreed to within the errors of their determination (typically 1\%). The exception is shown in Figs. \ref{periodograms} and \ref{lightcurves}. This is for a star known as HBC 338 \citep{hb}, an otherwise unremarkable T Tauri star in NGC 1333. Its position is $\alpha$ = 03:25:49.81, $\delta$ = +31:10:24.0 (J2000) and it is classified as a G8 to mid-K star with a V magnitude of 12.1 (but obviously variable). It was first noted as an H$\alpha$ emission star by \citet{s86}. \citet{hb} describe the H$\alpha$ emission as unusually broad and structured and classify it as ``tt". In today's nomenclature it would clearly qualify as a Classical T Tauri star since the strength and breadth of the H$\alpha$ emission indicate active accretion. Given its location on the sky, it is undoubtedly a member of NGC 1333, which is a small star-forming region in Perseus at a distance of around 300 pc \citep{b06}. 

\section{Results} 

Figs. \ref{periodograms} and \ref{lightcurves} demonstrate clearly that HBC 338 has significantly changed its rotation period with time. In the first two seasons the cluster was observed at Wesleyan, HBC 338 was found to have a period of 5.58 days and 5.57 days, respectively, an insignificant difference.  In the 2002-2003 season, the starÕs period is found to be 5.52, slightly but still not significantly less than its original period.\footnote{Typical accuracy of a period determination over a single observing season is about 1\%.} The light curve in this season is also now more scattered, possibly indicating the beginning of the break-up of the spot.  The fourth season's periodogram yields a primary period of 4.47 days, one day less than the star's period had been two years before, and significantly different.  In addition, this periodogram shows a second peak at 5.52 days, but of lesser power. The 2004-2005 season produced a period of 4.46 days, and is even more clearly defined than the previous year, showing that the star has settled into this new period and confirming its significance. As mentioned above, no other star has shown similar behavior and we have monitored more than a thousand T Tauri stars over time intervals as long as fifteen years at Wesleyan.

In an attempt to better understand what is going on with this star we have constructed a set of spot models that reproduce the main features of the light curves at each epoch. The models are of the simplest possible form since the data do not warrant anything more sophisticated. They consist of a two temperature photosphere with the cooler zone confined to a circular ``spot" of a given radius and latitude. A photospheric temperature of 4000 K and spot temperature of 3300 K were adopted based on the spectral type and typical spot temperatures found on other stars. Appropriate limb darkening coefficients were also adopted. Two fixed inclinations were chosen, 80$\degr$ and 90$\degr$, since the star must be viewed close to equator-on or it would not exhibit the light curve shapes we find (see below). Spot latitudes and radii were allowed to vary between 0$\degr$-90$\degr$ and 0$\degr$-45$\degr$ respectively. 

\begin{figure}
\plottwo{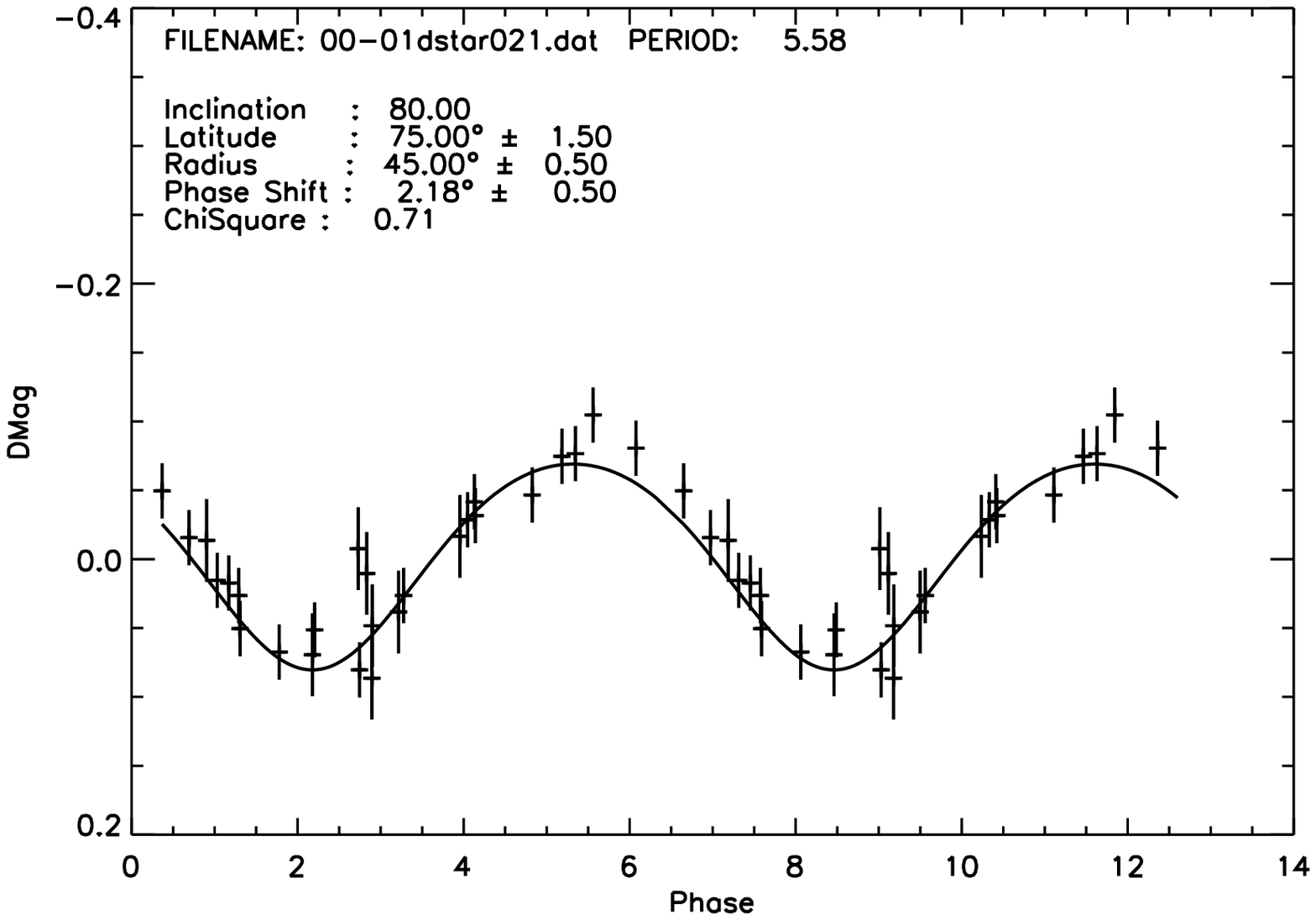}{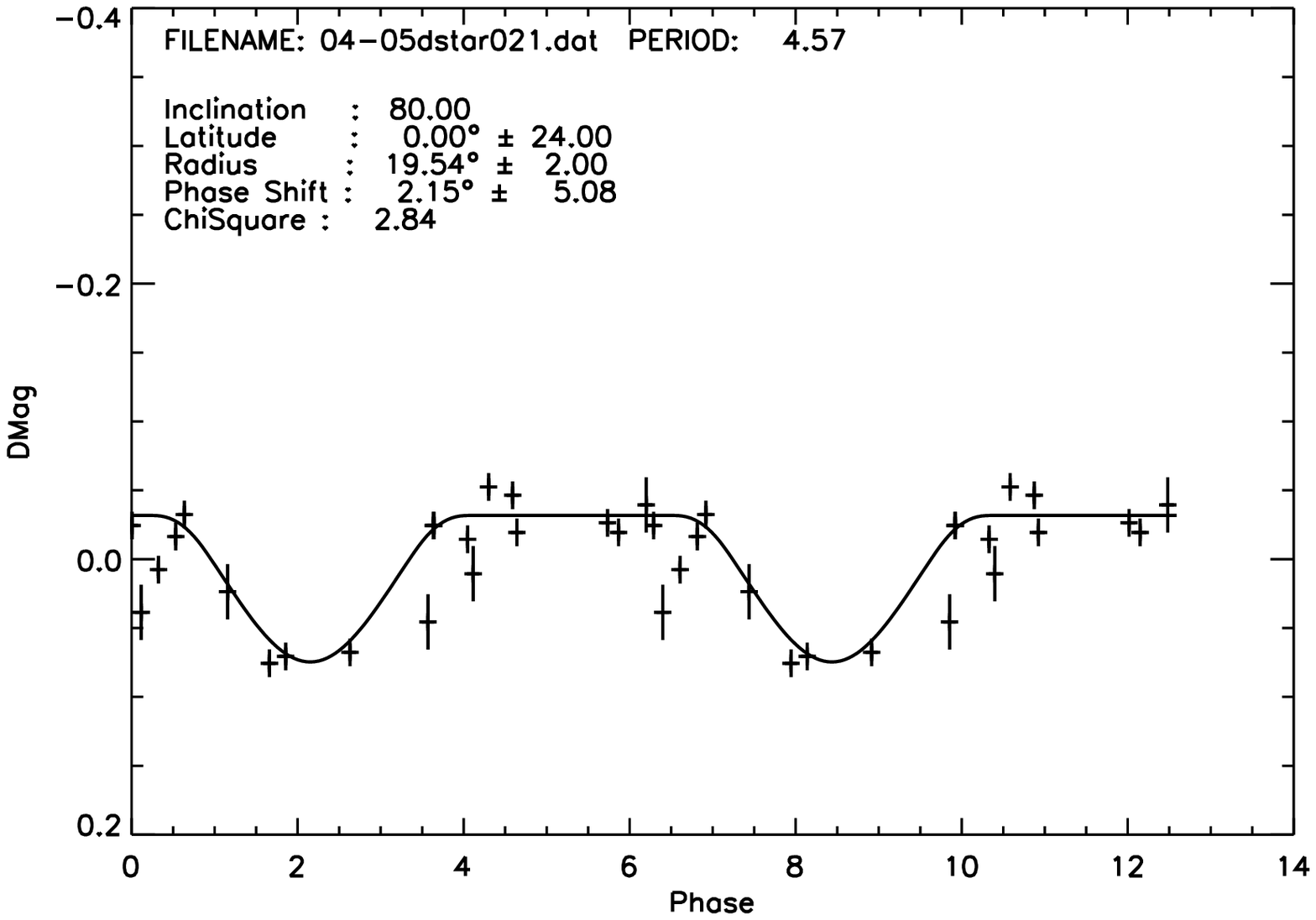}
\caption{Examples of the fit between the model light curves and the data in two seasons -- the first and last. In generating the displayed models, inclination was fixed at 80$\degr$ and spot radius-latitude parameter space was searched for the minimum value of chi-square. Clearly, an excellent fit is achieved for the first season and an adequate fit for the fifth season, where the data are more scattered, perhaps because of the influence of additional spots.} 
\label{fittedlightcurves}
\end{figure}

Best-fit light curves were chosen by a chi-squared minimization technique and two examples, the first and last seasons, are shown in Fig. \ref{fittedlightcurves}. The models respond to a feature which is quite obvious on the light curves themselves, namely that the shape of the light curve changes dramatically with period. In the first two years, when the period was longer, the light curve had a more peaked maximum and continuously variable form characteristic of a high latitude spot which straddles the rotation pole and, hence, is continuously visible. The latter two years, when the period was shorter, show a flat maximum, which is characteristic of an equatorial spot when the rotation axis is nearly equator-on. The flat maximum corresponds to the time that the spot is out of view of the observer on the opposite hemisphere of the star.

\begin{figure}
\plottwo{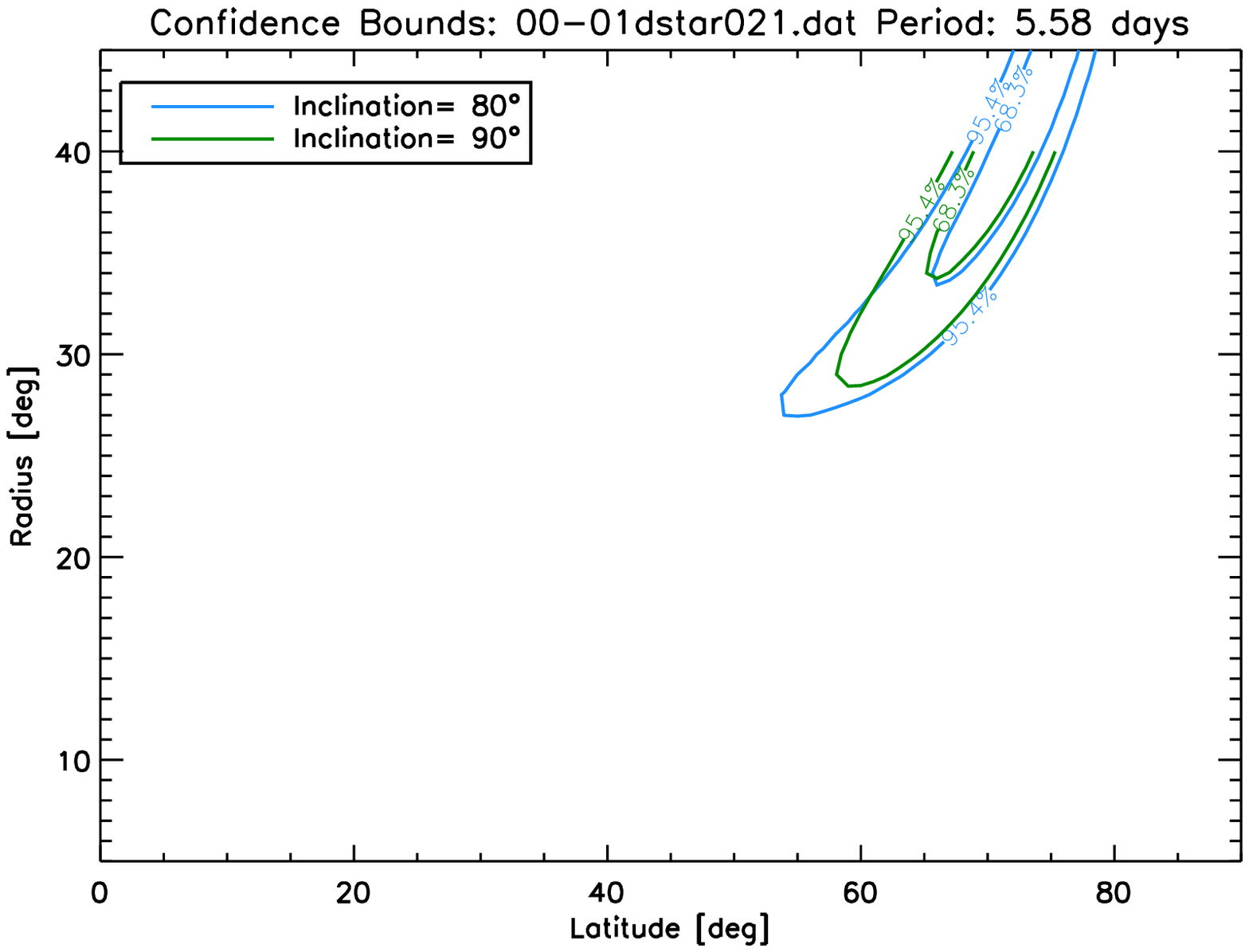}{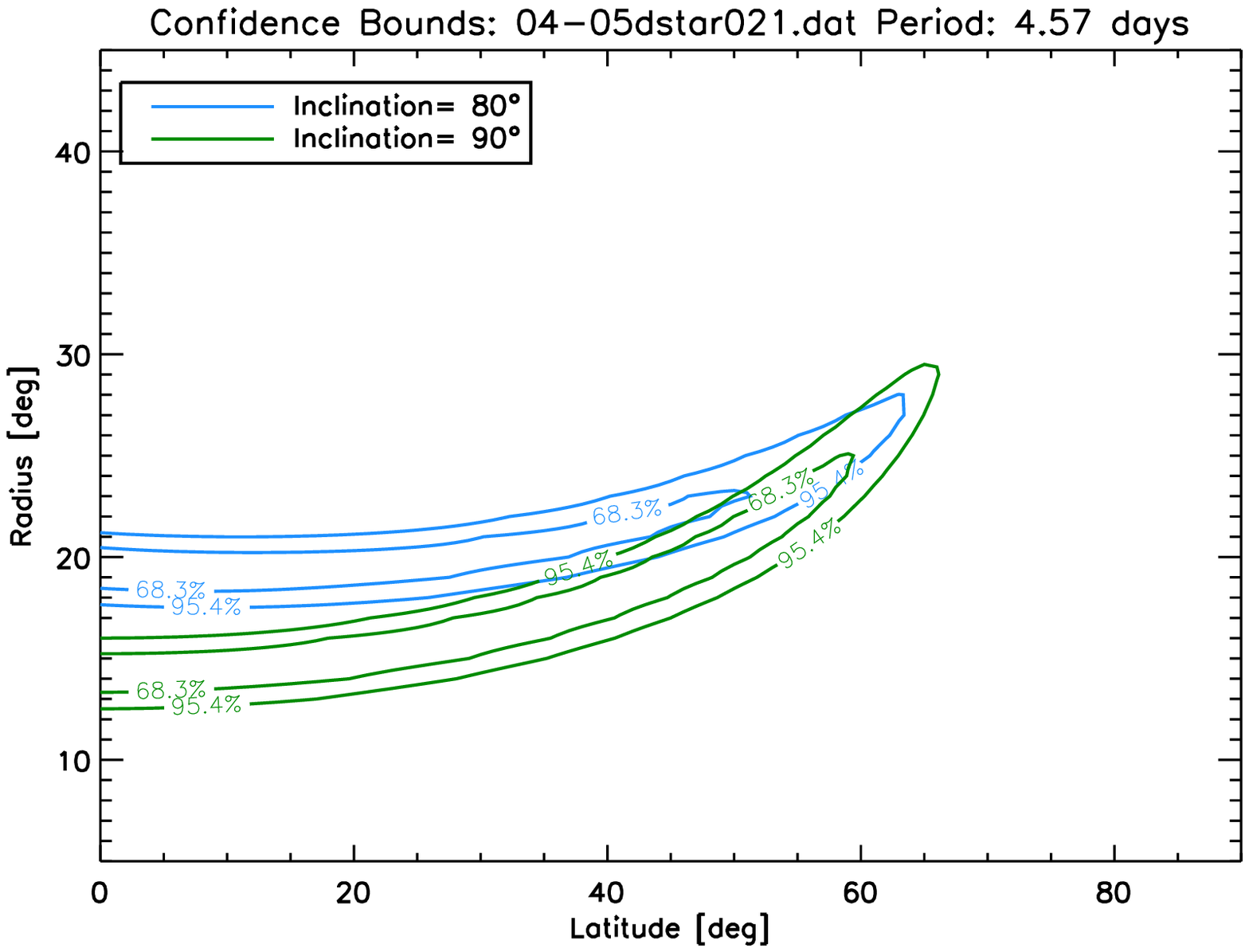}
\caption{Contours of equal confidence limit based on the chi square of various fits are shown on a spot radius-latitude grid. Two contour levels, corresponding to one and two sigma levels are shown for each of two inclination values, 80$\degr$ and 90$\degr$. We see that in the first season, when the rotation rate was slower, the data favor a high latitude spot of fairly large size. In the final season, when the rotation rate was faster, the data favor a lower latitude and smaller spot. The intermediate seasons yield intermediate results.} 
\label{confidencelimits}
\end{figure}

The fits we achieve for these simple models are remarkably good. A reduced chi-square of between 0.7 and 2.8 is achieved. To explore the range of parameters which would yield fits that are as good or nearly as good we calculated the reduced chi-square throughout parameter space and display the results in fig. \ref{confidencelimits}. Confidence limits are shown at the 68\% and 95\% levels for both the 80$\degr$ and 90$\degr$ inclinations. Clearly the results are not highly sensitive to inclination as long as the star is observed nearly equator on. If the inclination gets much smaller, however, it becomes impossible to obtain a good fit to the more recent spot light curves because of the flat portion.

What is most apparent and significant from this simple model fitting exercise is the point we made qualitatively above based on the shape of the light curves. Namely, the latitude of the dominant spot in the first two years must be high to explain the continuously varying (almost sinusoidal) shape. The models quantify this to be about 70$\degr$ in 2000/01 and 80$\degr$ in 2001/02 with an uncertainty of about 10$\degr$ in both years. By contrast, the latitude of the dominant spot in 2003/04 is much less constrained and could be equatorial, while by 2004/05 a polar spot is clearly ruled out.  

\section{Discussion and Summary}

Our observations demonstrate clearly, for the first time, that the rotation period measured for a T Tauri star can differ at different epochs by significant amounts, in this case about 20\%. Since it is unrealistic to think that the whole star could change its rotation rate on a time scale of one year, we see no reasonable alternative\footnote{For completeness we mention one unlikely alternative, namely that HBC 338 is a binary in which stars A and B have different rotation periods and have swapped overall brightness during the past five years. Since the mean magnitude of the system has remained unchanged over that time period this seems highly unlikely to us. However, without color and radial velocity information we cannot rule it out entirely.}\  to the hypothesis that we have detected differential rotation on the surface of the star. The amplitude of this effect is comparable to its amplitude on the Sun, whose period varies from 25.7 days at the equator to about 36 days at the pole \citep{b89}. 

Supporting this view is the fact that we notice a subtle but significant difference in the shapes of the light curves that correlates with their periods. During the first two years when the spot period was longer the light curve shapes are more peaked and best modeled as large, high latitude (70$\degr$-80$\degr$) spots which extend over the rotation pole and are, therefore, continuously visible from Earth. By contrast, in the last two years when the spot period was shorter, the light curves have flat maxima that are characteristic of spots close enough to the equator that they disappear entirely from view for some time. Spot models and chi-square minimization fitting techniques support these qualitative results.

Our conclusion is that the best interpretation of the data we have collected on this unusual star is the following. The star had a high latitude spot (or spot group) from 2001-03 which was gradually replaced by a more equatorial spot (or group) during 2003-05. The rotation period of the star depends on latitude in roughly the same manner and by about the same amount as observed for the Sun. Again, we note that perhaps the most surprising and significant thing about this star's behavior is that it is unique among the T Tauri stars we have monitored at VVO. The typical T Tauri star either rarely develops spots outside of a limited range of latitudes or rotates in a much more uniform way than does the Sun or HBC 338. Our guess, based on the typical light curve shapes, is that equatorial spots are very uncommon on T Tauri stars and that is why differential rotation is not usually detected. We expect that closer examination of spot curve shapes and continued monitoring over time scales of years or more will continue to reveal interesting details about the kinematics of these stars' surfaces.   

\acknowledgments

It is a pleasure to thank the many Wesleyan students
who carried out the observations. A.F.'s work was supported by the National Science
Foundation under Grant No. AST-0353997 to Wesleyan University,
supporting the Keck Northeast Astronomy Consortium (KNAC). S.D. was also supported by KNAC at a time when it was funded entirely by the member institutions. This material is based upon work supported by the National Aeronautics and Space Administration under Grant NNG05GO47G issued through the Origins of Solar Systems Program to W.H.

\clearpage

\clearpage

\clearpage

\clearpage

\end{document}